\documentstyle[12pt]{article}

\begin{document}
\baselineskip=24pt
\pagestyle{plain}

\title{The supergroup U(M/N)\\ with regard to electronic Hamiltonians}
\author{Ko Okumura\thanks{On leave from
 {\em Department of Physics, Faculty of Science and Technology,
Keio University, Yokohama 223, Japan}}\\
{\em Department of Physics, City College of}\\
{\em the City University of New York, New York, NY10031}}

\maketitle
\begin{abstract}
We study the U(M/N) supergroup keeping in mind its connection
with electronic Hamiltonians.
It is explicitly shown that the generators of the supergroup
U(N/N) can be expressed by Clifford operators or Fermi operators.
A multi-band supersymmetric electronic model is suggested.
\end{abstract}
\newpage

\section{Introduction}

Supersymmetry was discovered and has been developed
mainly in the context of high-energy physics\cite{SU}.
But recently the importance of supersymmetry in  electronic lattice
Hamiltonians has been noticed.
It is known that the t-J Hamiltonian becomes supersymmetric at 2t=J\cite{t-J}.
This U(1/2) supersymmetric model has been studied extensively partly
because it can be solved at least in one dimension\cite{t-J}.
A new electronic model for high-$T_c$ superconductivity
has been recently proposed by  Essler,  Korepin
and  Schoutens\cite{EKS}. This model is also solvable and has
a U(2/2) supersymmetry.
It was also found by the author in \cite{KO}(we call this `` I '' henceforward)
that the above two supersymmetries of the
models as well as the U(2) symmetry of the Heisenberg model
are manifest  when expressed in terms of U(2/2) operators.

With this in mind we study the U(M/N) group in the present paper.
As a straightforward generalization of I, we explicitly show that
U(N/N) group can be constructed  both from the Clifford operators
$\gamma^{\mu}$ and from the Fermi operators $c^{\pm}_{j\sigma}$.
The electronic Hamiltonians are usually described by Fermi operators.
On the one hand a transformation from Fermi operators to Clifford operators
is known\cite{MS}.
On the other hand we show that we can construct the U(N/N) supergroup
from $\gamma^{\mu}$.
Then it is natural that the U(N/N) supergroup  be realized directly
by the Fermi operators as we show.
A multi-band supersymmetric electronic Hamiltonian is also introduced.

The supergroup U(M/N) is defined as a transformation group on vectors
$\vec{x}$ on superspace V(M/N) which preserves the norm of these vectors.
Here we mean, by vector on V(M/N), a complex vector taking the form
\begin{equation}
\label{1-1}
\vec{x}=
\left(
\begin{array}{c}
x_1 \\ \vdots \\ x_M \\ x_{M+1} \\ \vdots \\ x_{M+N}
\end{array}
\right)
\equiv
\left(
\begin{array}{c}
\vec{x}_B \\ \vec{x}_F
\end{array}
\right)
\end{equation}
where the first M elements ($\vec{x}_B$) are ordinary complex numbers while
the last N elements ($\vec{x}_F$) complex Grassmann numbers.
Then, any element $G \in$ U(M/N) satisfies $G^+ G=1$.
The condition that $G\vec{x}$ must also be one of the elements
of V(M/N) requires the following form for $G$.
\begin{equation}
\label{1-2}
G=
\left(
\begin{array}{cc}
A & O \\ O & C
\end{array}
\right)
+
\left(
\begin{array}{cc}
O & D \\ E & O
\end{array}
\right)
\end{equation}
where $A$  ($C$) is a M$\times$M (N$\times$N) complex matrix
while $D$ ($E$) is a M$\times$N (N$\times$M) complex Grassmann matrix
whose matrix elements are complex Grassmann numbers.
Hereafter we call the block matrix like the first matrix in (\ref{1-2})
a {\em Bose block matrix} while the matrix like the second, a {\em
Fermi block matrix} (regardless of whether these matrix elements are
ordinary or Grassmann numbers).

As in ordinary Lie groups, let us introduce generators $H$ or
$B$ and $F$ in the following way.
\begin{equation}
\label{1-3}
G=e^{i\eta_j H_j}=e^{i(b_k B_k+f_l F_l)}
\end{equation}
where $b_k$'s ($f_l$'s) are {\em real} ({\em real Grassmann}) parameters
and $B_k$'s ($F_l$'s) are Bose (Fermi) block matrices.
Notice here that all the matrix elements of $B_j$ and $F_j$ are
just complex numbers.
Then the condition $G^+ G=1$ reduces to $B^{+}_{j} =B_j$ and
$F^{+}_{j} =F_j$, that is, the hermicity of the generators.
This condition makes the number of the independent generators or
the dimensionality of the group to be (M+N$)^2$, M$^2+$N$^2$ of which are
bosonic while 2MN are fermionic.
Here we mention that in order to form a group the following relations
are required.
\begin{equation}
\label{1-4}
\{F_j, F_k\}=f_{jkl}B_l\,,\quad
[F_j, B_k]=if_{jkl}F_l\,,\quad
[B_j, B_k]=if_{jkl}B_l
\end{equation}
where $f$'s are real numbers.
Hereafter we call commutation and anti-commutation relations like (\ref{1-4})
{\em graded commutation relations} in which  the anti-commutator
occurs only when both of the elements in the bracket are fermionic.

The (M$+$N)$^2$ generators $H$ can be (M$+$N)$\times$(M$+$N)
independent hermitian matrices
$T^{\alpha}$ ($\alpha=1,\ldots,$(M$+$N)$^2$) satisfying
\begin{equation}
\label{1-4'}
tr(T^{\alpha}T^{\beta})=\delta_{\alpha\beta}\,tr1,
\end{equation}
M$^2+$N$^2$ of which are Bose block matrices while 2MN are
Fermi block matrices.
$T^{\alpha}$ forms a closed superalgebra with
the graded commutator\footnote{A way of constructing these matrices
and explicit examples are given in Sec.~II}.

In this case any element of $G$ of U(M/N) takes the form
\begin{equation}
\label{1-5}
G=e^{i\eta_{\alpha}T^{\alpha}}
\end{equation}
where $\eta$'s are real parameters.

Another matrix representation of the U(M/N)
generators is given by $X^{ac}$;
\begin{equation}
\label{1-6}
(X^{ac})_{ij}=\delta_{ia}\delta_{jc}.
\end{equation}
where $a,c=1,\ldots,$M$+$N.
In other words $X$ is another basis of the U(M/N) generators, different from
$T$.
Indeed $T$ and $X$ are related by a linear transformation;
\begin{equation}
\label{1-7}
T^{\alpha}_{ij}=T^{\alpha}_{ac}X^{ac}_{ij}\quad\mbox{or}\quad
X^{ac}=\frac{1}{tr1}T^{\alpha}_{ca}T^{\alpha}.
\end{equation}
$X$ satisfies the following graded commutation rule which is a natural
generalization of the ordinary U(N) algebra.
\begin{equation}
\label{1-7'}
[X^{ac},X^{a'c'}]_{\pm}=X^{ac'}\delta_{a'c}\pm X^{a'c}\delta_{ac'}
\end{equation}
where $[$,$]_{\pm}$ denotes the graded commutator, that is,
commutator$(-)$ or anti-commutator(+)
 the latter occurring only if both $X$'s in the
bracket are Fermi block matrices.

In this case any element of $G$ of U(M/N) is given by
\begin{equation}
\label{1-8}
G=e^{i\zeta_{ac}X^{ac}}
\end{equation}
where complex parameter $\zeta$'s are defined as
$\zeta_{ac}=\eta_{\alpha}T^{\alpha}_{ac}$ with $a,c=1,\ldots,$M$+$N.

In the $X$ basis the algebra has the same form for any M and N while
in the $T$ basis it doesn't.
Notice here that in the (M$+$N)$\times$(M$+$N) matrix representation
any form $[M_1,M_2]$ or $\{M_1,M_2\}$ where
$M_1,M_2 \in T^{\alpha}$ or $X^{ac}$ is a (M$+$N)$\times$(M$+$N) matrix
so that it can be expressed in terms of a linear combination of
$T^{\alpha}$ or $X^{ac}$. Thus we can form
U(M$+$N),U(M$+$N$-$1/1),$ $\ldots,$ $U(1/M$+$N$-$1) from
$T^{\alpha}$ or $X^{ac}$.

\setcounter{equation}{0}
\section{The construction of U(N/N) generators from
Clifford or Fermi operators}

In this section we explicitly show that the matrix representation
of the U(N/N)\footnote{We can get U(M/N) generators (M$\neq$N) in terms of
these operators by applying appropriate projection operators
as the examples are shown in Sec.~III.}
generators $T$ and $X$ can be realized by the $\hat{T}$ and $\hat{X}$
operators which are constructed from the Clifford($\gamma$)
and the Fermi($c$ and $c^+$) {\em operators} respectively.
These $\hat{T}$ and $\hat{X}$ operators are related through
\begin{equation}
\label{2-0}
\hat{X}^{ac}=\frac{1}{tr1}T^{\alpha}_{ca}\hat{T}^{\alpha}
\quad \mbox{or} \quad
\hat{T}^{\alpha}=T^{\alpha}_{ac}\hat{X}^{ac}
\end{equation}
where $T^{\alpha}_{ac}$ is the $(a,c)$ element of the matrix
$T^{\alpha}$.
Eqs.~(\ref{2-0}) implies the following relation\cite{MS};
\begin{equation}
\label{2-0'}
\gamma^{2l-1}=c^l+(c^l)^+ \, , \quad
\gamma^{2l}=-i(c^l-(c^l)^+)
\end{equation}
which is the transformation between the $\gamma$ (O(N)) operators
and the Fermi (U(N)) operators.

\subsection{The construction of $T^{\alpha}$ from Clifford operators}

{}From the   $\gamma^{\mu}$ operators which form
the 2$D$-dimensional Clifford algebra
\begin{equation}
\label{2-1}
\{\gamma^{\mu},\gamma^{\nu}\}=2\delta_{\mu\nu}\quad
(\mu,\nu=1,\ldots,2D),
\end{equation}
we can construct through multiplication of $\gamma$'s,
$2^{2D}$ independent hermitian operators
$\hat{T}^{\alpha}$ $(\alpha=1,\ldots,2^{2D})$  which satisfy
\begin{equation}
\label{2-2}
tr(\hat{T}^{\alpha}\hat{T}^{\beta})=\delta_{\alpha\beta}tr1.
\end{equation}
We give an explicit way for this construction in the following table.
  \\

\begin{tabular}{|c|c||c|c|} \hline
\label{t1}
$\hat{T}^{\alpha}$ & \# of indep. operators &
$\hat{T}^{\alpha}$ $(D=1)$ & $\hat{T}^{\alpha}$ $(D=2)$ \\ \hline
1  & $_{2D}C_{0}=1$ & 1 & 1 \\
$\gamma^{\mu}$ &
$_{2D}C_{1}=2D$ &
$\gamma^{\mu}$ & $\gamma^{\mu}$ \\
$(\gamma^{\mu_1}\gamma^{\mu_2}-\gamma^{\mu_2}\gamma^{\mu_1})/(2i)$ &
$_{2D}C_{2}$ &
-- & $\sigma^{\mu\nu}$ \\
$\langle\langle\gamma^{\mu_1}\gamma^{\mu_2}\gamma^{\mu_3}
\rangle\rangle$ &
$_{2D}C_{3}$ &
-- & $i\gamma^{5}\gamma^{\mu}$ \\
$\vdots$ & $\vdots$ & $\vdots$ & $\vdots$ \\
$\langle\langle\gamma^{\mu_1}\cdots\gamma^{\mu_l}\rangle\rangle$ &
$_{2D}C_{l}$ &
-- & -- \\
$\vdots$ & $\vdots$ & $\vdots$ & $\vdots$ \\
$\langle\langle\gamma^{\mu_1}\cdots\gamma^{\mu_{2D}}\rangle\rangle$ &
$_{2D}C_{2D}=1$ &
$\sigma^{\mu\nu}$ & $\gamma^5$ \\ \hline
total \# of indep. operators &
$ \sum_{l=0}^{2D} {}_{2D}C_{l}=2^{2D}$ & 4 & 16 \\ \hline
\end{tabular}
\\

where $_{n}C_{r}=\frac{n!}{(n-r)!\,r!}$ and
\begin{equation}
\label{2-2'}
\langle\langle \gamma^{\mu_1}\cdots\gamma^{\mu_{l}}\rangle\rangle
=\left\{ \begin{array}{ll}
-\gamma^{\mu_1}\cdots\gamma^{\mu_{l}} & l=4m, 4m+1 \\
-i\gamma^{\mu_1}\cdots\gamma^{\mu_{l}} &   l=4m+2, 4m+3
\end{array} \right.
\end{equation}
where m is an integer.
Note that the operators in the even (odd) lines are Bose (Fermi) operators and
both are $2^{2D-1}$ in number
$( \sum_{l=0}^{D}  {}_{2D}C_{2l}=
\sum_{l=0}^{D-1}  {}_{2D}C_{2l+1}=2^{2D-1}).$

In this construction it is easy to see that (\ref{2-2}) actually hold.
So $\hat{T}^{\alpha}$ is an abstract representation of the generators
of U(N/N) with N=$2^{D-1}$.
In order to get a 2N$\times$2N matrix representation or a defining
representation of
$T^{\alpha}$ from $\hat{T}^{\alpha}$, we  first write down $2D$ 2N$\times$2N
independent Fermi block matrices and identify them with $\gamma^{\mu}$
$(\mu=1,\ldots,2D)$. Then through multiplication of these 2N$\times$2N
$\gamma^{\mu}$ matrices we get the other $T^{\alpha}\,$'s.
An example for $D=2$ case is given in Sec.~III.

Notice here that the $\hat{T}^{\alpha}$ thus obtained can form
U(2N) etc.\ other than U(N/N) with appropriate redefinition of the
Fermi and Bose operators.
Non the less if we introduce local $\gamma^{\mu}_{j}$ $(j=1,\ldots,L)$
at each {\em site} and require anti-commutation rule among different
site operators, that is,
$\{\gamma^{\mu}_{j},\gamma^{\nu}_{k}\}=2\delta_{jk}\delta_{\mu\nu}$
and construct $T^{\alpha}_{j}$ then the global
$\tilde{T}^{\alpha}$ $(=\sum_{j=1}^{L}T^{\alpha}_{j})$
forms only U(N/N).

\subsection{The construction of $X^{ac}$ from Fermi operators}

{}From $2n$ Fermi operators $(c^l)^{\pm}$ where $(c^l)^-=c^l$
with $l=1,\ldots,2n$\footnote{$l$ tacitly denotes
{\em band}- and {\em spin}-indices.} which satisfy
\begin{equation}
\label{2-3}
\{c^l,(c^{l})^+\}=\delta_{ll'}
\end{equation}
we can construct $2^{2n}\times2^{2n}$ independent real operators
$X^{ac}$ $(a,c=1,\ldots,2^{2n})$ which satisfy
\begin{equation}
\label{2-4}
[\hat{X}^{ac},\hat{X}^{a'c'}]_{\pm}=
\hat{X}^{ac'}\delta_{a'c}\pm \hat{X}^{a'c}\delta_{ac'}.
\end{equation}

Let us first introduce vacuum state $|0\rangle$ which is annihilated
by $c^l$; $c^l|0\rangle=0$.
Then  all the possible independent state are as follows.
\\

\begin{tabular}{|c|c|} \hline
\label{t2}
Type of states & \# of indep. states \\ \hline
$|0\rangle$ & $_{2n}C_{0}=1$ \\
$(c^{l_1})^+|0\rangle$ & $_{2n}C_{1}=2n$ \\
$(c^{l_1})^+(c^{l_2})^+|0\rangle$ & $_{2n}C_{2}$ \\
$\vdots$ & $\vdots$ \\
$(c^{l_1})^+\cdots (c^{l_r})^+|0\rangle$ & $_{2n}C_{r}$ \\
$\vdots$ & $\vdots$ \\
$(c^{l_1})^+\cdots (c^{l_2n})^+|0\rangle$ & $_{2n}C_{2n}=1$ \\ \hline
total \# of indep. states &
$\sum_{r=0}^{2n} {}_{2n}C_{r} =2^{2n}$ \\ \hline
\end{tabular}
\\

Now we can introduce $2^{2n}\times2^{2n}$ independent operators
$\hat{X}^{ac}$;
\begin{equation}
\label{2-5}
\hat{X}^{ac}=|a\rangle \langle c | \quad (a,c=1,\ldots,2^{2n})
\end{equation}
It is easy to check that (\ref{2-4}) holds.
Noting that
\begin{equation}
|0\rangle \langle 0| = \,\, :e^{-(c^l)^+ c^l}: \,\,
= \,\, :\prod_l (1-(c^l)^+ c^l): \,\, ,
\end{equation}
Eq.(\ref{2-0}) may be derived directly.
$X^{ac}$ is called Fermi (Bose) type if the total number of
Fermi operators in $|a\rangle \langle c |$ is odd (even).
Notice here that the number of independent Bose and Fermi
operators is $2^{4n-1}$ each.
So $\hat{X}^{ac}$ is an abstract representation of the generators
of U(N/N) with N=$2^{2n-1}$.
The 2N$\times$2N matrix representation
is clearly given by (\ref{1-6}).

Notice here that the $\hat{X}^{ac}$ thus obtained can form
U(2N) etc.\  other than U(N/N) with appropriate redefinition of
the Bose and Fermi
operators.
If we introduce local $c^{l}_{j}$
 at each site $(j=1,\ldots,L)$, require  ordinary fermion
anti-commutation relation
\begin{equation}
\label{2-6}
\{c^{l}_{j},(c^{l'}_{j'})^+\}=\delta_{jj'}\delta_{ll'}
\end{equation}
on it and define local $X^{ac}_{j}=|a_j\rangle \langle c_j | $
then the $X^{ac}_{j}$ satisfies
\begin{equation}
\label{2-7}
[X^{ac}_j, X^{a'c'}_{j'}]_{\pm}
=\delta_{jj'}(X^{ac'}_{j}\delta_{a'c}\pm X^{a'c}_{j}\delta_{ac'}).
\end{equation}
Here we have introduced the {\em local} vacuum $|0\rangle_j$
which is annihilated by $c^{l}_{j}$ and
$|a_j\rangle =(c^{l_1})^+\cdots (c^{l_k})^+|0\rangle_j$.
In this case the vacuum is defined as
\begin{equation}
\label{2-8}
|0\rangle=\otimes_j |0\rangle_j
\end{equation}
which satisfies $c^{l}_{j}|0\rangle =0$.
Now if we define the global $\tilde{X}^{ac}=\sum_{j=1}^{L}X^{ac}_{j}$,
then $\tilde{X}^{ac}$ forms only U(N/N).

\setcounter{equation}{0}
\section{Examples from condensed matter physics}

Let us consider a one-band electronic lattice Hamiltonian
like Hubbard model which has L lattice points\cite{KO}.
Then the Hilbert space is spanned by the states of the form
\begin{equation}
\label{3-1}
|a_1\rangle\otimes\cdots\otimes|a_j\rangle\otimes\cdots\otimes|a_L\rangle
\end{equation}
where $|a_j\rangle$ is $|1_j\rangle=|0\rangle_j$,
$|2_j\rangle=c^{+}_{j\uparrow}c^{+}_{j\downarrow}|0\rangle_j$,
$|3_j\rangle=c^{+}_{j\uparrow}|0\rangle_j$, or
$|4_j\rangle=c^{+}_{j\downarrow}|0\rangle_j$.
Here $|0\rangle_j$ is defined as $c_{j\sigma}|0\rangle_j=0$
and the vacuum $|0\rangle=\otimes|0\rangle_j$ satisfies
$c_{j\sigma}|0\rangle=0$.
Then we can introduce 16 $X^{ac}_{j}$ operators at each site
and construct the global $\tilde{X}^{ac}$ which forms U(2/2).

Next we consider the restricted Hilbert space
$P|a_1\rangle\otimes\cdots\otimes|a_L\rangle$
in which doubly occupied states $|2_j\rangle$ are missing.
Here we have introduced the Gutzwiller projection operator $P$
(defined in I).
On this space $X^{ac}_{j}$ becomes $PX^{ac}_{j}P$ (or $X^{ac}_{j}$
with only $a,c=1,3,4$). This restricted $X^{ac}_{j}$ is the
Hubbard projection operator and $\tilde{X}^{ac}$ forms U(1/2).

Further, consider other restricted spaces
$P_{F}|a_1\rangle\otimes\cdots\otimes|a_L\rangle$
and
$P_{B}|a_1\rangle\otimes\cdots\otimes|a_L\rangle$
in which $|1_j\rangle$, $|2_j\rangle$ and
$|3_j\rangle$, $|4_j\rangle$ states are missing respectively.
Here $P_{F}$ and $P_{B}$ are also defined in Eqs.(2.12) and (2.13) of I.
It is clear that both $P_{F}\tilde{X}^{ac}P_{F}$ and $P_{B}\tilde{X}^{ac}P_{B}$
satisfy the U(2) algebra. Notice here that in these cases all the Fermi
operators
of $X^{ac}_{j}$ are missing and thus the superalgebra reduces to
an ordinary algebra.

In terms of $\hat{T}^{\alpha}$ or $\gamma$ operators, we first define
$\gamma$ through Eq.(2.1) of I and then the 16 $\hat{T}^{\alpha}_j$ become
$1, \gamma^{\mu}_{j}, \sigma^{\mu\nu}_{j},
i\gamma^{5}\gamma^{\mu}_{j}, \gamma^{5}_{j}$.
In this case the 16 $T^{\alpha}\,$'s
$(=(1, \Gamma^{\mu}, \Sigma^{\mu\nu}, iA^{\mu}, \Gamma^{5}))$
are explicitly given by (2.5) of I and
the U(2/2) algebra becomes
\begin{eqnarray}
\label{3-3}
\mbox{[}\Sigma^{\mu\nu},\Sigma^{\rho\tau}]
& = &
4i(\delta_{\mu\tau}\Sigma^{\nu\rho}+\delta_{\nu\rho}\Sigma^{\mu\tau}
-\delta_{\mu\rho}\Sigma^{\nu\tau}-\delta_{\nu\tau}\Sigma^{\mu\rho}),
\quad
\mbox{[} \Sigma^{\mu\nu}, \Gamma^{5} ]  =
\mbox{[}\Gamma^{5},\Gamma^{5} ]  =  0
\nonumber \\
\{\Gamma^{\mu},\Gamma^{\nu}\} & = &
\{iA^{\mu},iA^{\nu}\}  = 2\delta_{\mu\nu},
\quad
\{\Gamma^{\mu},iA^{\nu}\}  =  \varepsilon^{\mu\nu\rho\tau}\Sigma^{\rho\tau}
\nonumber \\
\mbox{[}\Gamma^{\mu},\Sigma^{\rho\tau}] & = &
2i(\delta_{\mu\rho}\Gamma^{\tau}-\delta_{\mu\tau}\Gamma^{\rho}),
\quad
\mbox{[}\Gamma^{\mu},\Gamma^{5}] =  2i(iA^{\mu})
\nonumber \\
\mbox{[}iA^{\mu},\Sigma^{\rho\tau}] & = &
2i(\delta_{\mu\rho}iA^{\tau}-\delta_{\mu\tau}iA^{\rho}),
\quad
\mbox{[}iA^{\mu},\Gamma^{5}] =  -2i(\Gamma^{\mu})
\end{eqnarray}

Next we consider a 2-band electronic system where $d$ and $f$ electrons are
described by canonical Fermi operator $d^{\pm}_{j\sigma}$
and $f^{\pm}_{j\sigma}$ which satisfy the usual anti-commutation rule.
Introducing 16 independent states following the table in Sec.~II.~2
we define 256 independent $X^{ac}_{j}\,$'s at each site.
Then the global $\tilde{X}^{ac}$ form U(8/8) and appropriate projection
operators reduce U(8/8) to other U(M/N).

In terms of the $\gamma$ operator we introduce 8 $\gamma^{\mu}$
matrix as real and imaginary parts of $d^{\pm}$ and
$f^{\pm}$ (see (\ref{2-0'})) and construct the $T$ operators.

Finally we would like to mention that
the U(M/N) symmetry of the model manifests itself if we rewrite
Hamiltonians in terms of the $X$ operators\cite{KO}.
As a straightforward generalization of the result given in I
the  electronic lattice Hamiltonian given by
\begin{equation}
\label{r}
H=-t\sum_{<jk>}\sum_{ac}X^{ac}_j X^{ca}_k (-1)^{F(c)}
\end{equation}
for $n$-band electronic system
$(a,c = 1,2,\ldots , 2^{2n-1})$
will be interesting.
This model has a U($2^{2n-1}$/$2^{2n-1}$) symmetry.

\section*{Acknowledgment}

I would like to thank  Professor Bunji Sakita
 for  suggestions, discussions, and encouragement
throughout this work.
I also thank the Rotary Foundation for a scholarship.
This work was supported by NSF grant PHY90$-$20495 and
PSCBHE grant 6$-$63351.


\begin{thebibliography}{99}
\bibitem{SU}
For general reference to supersymmetry, see, for example, the following.
P.\ G.\ O.\ Freund and I.\ Kaplansky, J.\ Math.\ Phys.\ {\bf 17}, 228 (1976);
V.\ Rittenberg, {\em A Guide to Lie Superalgebras}, in Supergravities in
diverse dimension Vol.\ 1, 64 (North-Holland/World Scientific 1989);
P.\ G.\ O.\ Freund, {\em Introduction to Supersymmetry}
(Cambridge University Press 1986)
\bibitem{t-J}
P.\ B.\ Wiegmann, Phys.\ Rev.\ Lett.\ {\bf 60}, 821 (1988);
D.\ Forster, Phys.\ Rev.\ Lett.\ {\bf 63}, 2140 (1989);
S.\ Sarker, J.\ Phys.\ {\bf A23}, L409 (1990);
S.\ Sarker, J.\ Phys.\ {\bf A24}, 1137 (1991);
P.\ A.\ Bares, G.\ Blatter and M.\ Ogata, Phys.\ Rev.\ {\bf 44}, 130 (1991).
\bibitem{EKS}
F.\ H.\ L.\ Essler, V.\ E.\ Korepin and K.\ Schoutens,
Phys.\ Rev.\ Lett.\ {\bf 68}\ 2960 (1992) and
{\em Electronic model for superconductivity}, preprint ITP-SB-92-20.
\bibitem{KO}K.\ Okumura, {\em Various condensed matter Hamiltonians
in terms of U(2/2) operators and their symmetry structures},
Mod.\ Phys.\ Lett.\ {\bf B} (to be published).
\bibitem{MS} R.\ N.\ Mohapatra and B.\ Sakita, Phys.\ Rev.\ {\bf D21},
1062 (1980).
\end{thebibliography}
\end{document}